\documentstyle[12pt]{article}
\newfont{\bbd}{msbm10 scaled\magstep1}

\begin{document}
\thispagestyle{empty}

\def\ve#1{\mid #1\rangle}
\def\vc#1{\langle #1\mid}

\newcommand{\p}[1]{(\ref{#1})}
\newcommand{\be}{\begin{equation}}
\newcommand{\ee}{\end{equation}}
\newcommand{\sect}[1]{\setcounter{equation}{0}\section{#1}}


\newcommand{\vs}[1]{\rule[- #1 mm]{0mm}{#1 mm}}
\newcommand{\hs}[1]{\hspace{#1mm}}
\newcommand{\mb}[1]{\hs{5}\mbox{#1}\hs{5}}
\newcommand{\Db}{{\overline D}}
\newcommand{\bea}{\begin{eqnarray}}

\newcommand{\eea}{\end{eqnarray}}
\newcommand{\wt}[1]{\widetilde{#1}}
\newcommand{\und}[1]{\underline{#1}}
\newcommand{\ov}[1]{\overline{#1}}
\newcommand{\sm}[2]{\frac{\mbox{\footnotesize #1}\vs{-2}}
           {\vs{-2}\mbox{\footnotesize #2}}}
\newcommand{\prt}{\partial}
\newcommand{\eps}{\epsilon}

\newcommand{\R}{\mbox{\rule{0.2mm}{2.8mm}\hspace{-1.5mm} R}}
\newcommand{\Z}{Z\hspace{-2mm}Z}

\newcommand{\cd}{{\cal D}}
\newcommand{\cg}{{\cal G}}
\newcommand{\ck}{{\cal K}}
\newcommand{\cw}{{\cal W}}

\newcommand{\vj}{\vec{J}}
\newcommand{\vl}{\vec{\lambda}}
\newcommand{\vz}{\vec{\sigma}}
\newcommand{\vt}{\vec{\tau}}
\newcommand{\vw}{\vec{W}}
\newcommand{\poiss}{\stackrel{\otimes}{,}}

\def\l#1#2{\raisebox{.2ex}{$\displaystyle
  \mathop{#1}^{{\scriptstyle #2}\rightarrow}$}}
\def\r#1#2{\raisebox{.2ex}{$\displaystyle
 \mathop{#1}^{\leftarrow {\scriptstyle #2}}$}}



\renewcommand{\thefootnote}{\fnsymbol{footnote}}
\newpage
\setcounter{page}{0}
\pagestyle{empty}
\begin{flushright}
{June 2002}\\
{LPENSL--TH--05/2002}
\end{flushright}
\vfill

\begin{center}
{\LARGE {\bf Recursion operators of the}}\\[0.3cm]
{\LARGE {\bf N=2 supersymmetric unconstrained}}\\[0.3cm]
{\LARGE {\bf matrix GNLS hierarchies}}\\[1cm]

{}~

{\large F. Delduc$^{a,1}$ and A.S. Sorin$^{b,2}$}
{}~\\
\quad \\
{\em {~$~^{(a)}$ Laboratoire de Physique$^\dagger$,
Groupe de Physique Th\'eorique,}}\\
{\em ENS Lyon, 46 All\'ee d'Italie, 69364 Lyon, France}\\[10pt]
{\em {~$~^{(b)}$ Bogoliubov Laboratory of Theoretical Physics,}}\\
{\em {Joint Institute for Nuclear Research,}}\\
{\em 141980 Dubna, Moscow Region, Russia}~\quad\\

\end{center}

\vfill

{}~

\centerline{{\bf Abstract}} \noindent A super--algebraic
formulation of the $N=2$ supersymmetric unconstrained matrix
$(k|n,m)$--MGNLS hierarchies (nlin.SI/0201026) is established.
Recursion operators, fermionic and bosonic symmetries as well as
their superalgebra are constructed for these hierarchies.

{}~

{}~



{}~

{}~

\vfill
{\em \noindent
1) E-Mail: francois.delduc@ens-lyon.fr\\
2) E-Mail: sorin@thsun1.jinr.ru }\\
$\dagger$) UMR 5672 du CNRS, associ\'ee \`a l'Ecole Normale Sup\'erieure de
Lyon.
\newpage
\pagestyle{plain}
\renewcommand{\thefootnote}{\arabic{footnote}}
\setcounter{footnote}{0}

\section{Introduction}

The $N=2$ supersymmetric unconstrained matrix
$(k|n,m)$--Generalized Nonlinear Sch\"odinger ($(k|n,m)$--MGNLS)
hierarchies were proposed in \cite{ks1} by exhibiting the
corresponding {\it matrix pseudo--differential} Lax--pair
representation in terms of {\it $N=2$ unconstrained superfields}
for the bosonic isospectral flows. These hierarchies generalize
and contain as limiting cases many other interesting $N=2$
supersymmetric hierarchies discussed in the literature. When
matrix entries are chiral and antichiral $N=2$ superfields, these
hierarchies reproduce the $N=2$ chiral matrix $(k|n,m)$-GNLS
hierarchies \cite{bks1,bks2}, and in turn the latter coincide with
the $N=2$ GNLS hierarchies of references \cite{bks,bs} in the
scalar case $k=1$. When matrix entries are unconstrained $N=2$
superfields and $k=1$, these hierarchies are equivalent to the
$N=2$ supersymmetric multicomponent hierarchies \cite{pop}. The
bosonic limit of the $N=2$ unconstrained $(k|0,m)$--MGNLS
hierarchy reproduces the bosonic matrix NLS equation elaborated in
\cite{fk} via the $gl(2k+m)/(gl(2k)\times gl(m))$--coset
construction. The $N=2$ $(1|1,0)$--MGNLS hierarchy is related to
one of three different existing $N=2$ supersymmetric KdV
hierarchies -- the $N=2$ ${\alpha}=1$ KdV hierarchy -- by a
reduction \cite{pop,ks1,ks2}.

Apart from the Lax--pair representation for the isospectral
bosonic flows of the $N=2$ unconstrained $(k|n,m)$--MGNLS
hierarchies, at present we do not know other quantities and/or
data (if any) which could characterize their integrable structure,
like, e.g. their super--algebraic formulation, bosonic and
fermionic symmetries, Hamiltonian structures, recursion operators,
etc. (although part of these are known for some of
above--mentioned limiting cases).

The present letter addresses these problems. We obtain a
super--algebraic formulation of the $N=2$ unconstrained
$(k|n,m)$--MGNLS hierarchies. Using it and the superalgebraic
methods developed in refs. \cite{dm1,dm2,dfg,dg,mm,adr,adrz} and
especially \cite{agnpz} we derive the superalgebra of fermionic
and bosonic symmetries as well as the recursion operators for
these hierarchies.

The letter is organized as follows. In Section 2.1 we present a
short summary of the pseudo--differential Lax--pair approach to
the $N=2$ unconstrained $(k|n,m)$--MGNLS hierarchies. In Section
2.2 we rewrite the corresponding spectral equation in a local
matrix form and establish its super--algebraic structure which is
then used in Section 2.3 and 2.4 to derive the superalgebra of the
symmetries and the recursion operators of the hierarchy, respectively.
In Section 2.5 we discuss supersymmetry and locality of the isospectral
flows. In Section 3 we summarize our results and discuss open problems.

\section{The $N=2$ unconstrained $(k|n,m)$--MGNLS hierarchies}

\subsection{Pseudo--differential Lax pair representation}

The Lax--pair representation for the bosonic flows
of the $N=2$ supersymmetric unconstrained $(k|n,m)$--MGNLS hierarchies is
\cite{ks1}
\begin{eqnarray}
{\textstyle{\partial\over\partial t_p}}L =[ A_p , L],\quad
L = I\partial + F D{\overline D}\partial^{-1} {\overline F},
\quad  A_p = (L^p)_{\geq 0} + res(L^p), \quad p \in \hbox{\bbd N}
\label{suplax}
\end{eqnarray}
where the subscript $\geq 0$ denotes the sum of purely
differential and constant parts of the operator $L^p$, and
$res(L^p)$ is its $N=2$ supersymmetric residue, i.e. the
coefficient of $[D,\overline D]{\partial}^{-1}$. Here, $F\equiv
F_{Aa}(Z)$ and ${\overline F}\equiv {\overline F}_{aA}(Z)$
($A,B=1,\ldots, k$; $a,b=1,\ldots , n+m$) are rectangular matrices
which entries are unconstrained $N=2$ superfields, $I$ is the
unity matrix, $I_{AB}\equiv {\delta}_{AB}$, and the matrix product
is implied, for example $(F\overline F)_{AB} \equiv \sum_a
F_{Aa}\overline F_{aB}$. The matrix entries are Grassmann even
superfields for $a=1,\ldots ,n$ and Grassmann odd superfields for
$a=n+1,\ldots , n+m$. Thus, fields do not commute, but rather
satisfy $F_{Aa}{\overline F}_{bB}=(-1)^{d_{a}{\overline d}_{b}}
{\overline F}_{bB}F_{Aa}$ where $d_{a}$ and ${\overline d}_{b}$
are the Grassmann parities of the matrix elements $F_{Aa}$ and
${\overline F}_{bB}$, respectively, $d_{a}=1$ $(d_{a}=0)$ for odd
(even) entries. Fields depend on the coordinates
$Z=(z,\theta,\overline\theta)$ of $N=2$ superspace. The volume
element in superspace is $dZ \equiv dz d \theta d
\overline\theta$. Finally, $D,{\overline D}$ are the $N=2$
supersymmetric fermionic covariant derivatives
\begin{eqnarray}
D=\frac{\partial}{\partial\theta}
 -\frac{1}{2}\overline\theta\frac{\partial}{\partial z}, \quad
{\overline D}=\frac{\partial}{\partial\overline\theta}
 -\frac{1}{2}\theta\frac{\partial}{\partial z}, \quad
D^{2}={\overline D}^{2}=0, \quad
\left\{ D,{\overline D} \right\}= -\frac{\partial}{\partial z}
\equiv -{\partial}.
\label{DD}
\end{eqnarray}
The algebra of the flows in \p{suplax} is abelian
\begin{eqnarray}
[{\textstyle{\partial\over\partial t_m}},
{\textstyle{\partial\over\partial t_n}}]=0 .\label{alg}
\end{eqnarray}
The Lax pair representation \p{suplax} may be seen as the
integrability condition for the following linear system:
\begin{eqnarray}
L\psi_1  &=& {\lambda} \psi_1, \label{spectrstart} \\
{\textstyle{\partial\over\partial t_p}}\psi_1 &=& A_p \psi_1
\label{spectr}
\end{eqnarray}
where $\lambda$ is the spectral parameter and the eigenfunction $\psi_1$
is the Baker-Akhiezer function of the hierarchy.

\subsection{Matrix formulation of the spectral equation}

Let us rewrite the spectral equation \p{spectrstart} in a matrix
form in $N=2$ superspace
\begin{eqnarray}
{\cal L}{\Psi}=0, \quad
{\overline {\cal L}}{\Psi}=0, \quad
\Psi =
\left(\begin{array}{c}
\psi_1 \\
\psi_2 \\
\psi_3 \\
\psi_4 \\
\psi_5
\end{array}\right)
\label{spectrferm}
\end{eqnarray}
with two $N=2$ odd Lax operators ${\cal L}$ and ${\overline {\cal
L}}$
\begin{eqnarray}
{\cal L}= D + A_{\theta}, \quad
{\overline {\cal L}}= {\overline D} + A_{\overline \theta}
\label{LN=2}
\end{eqnarray}
whose odd connections $A_{\theta}$ and $A_{\overline \theta}$ are
restricted to be local functionals of the original superfield
matrices $F$ and $\overline F$ and their $\{D,{\overline
D}\}$--derivatives. One finds that the eigenvalue equation
(\ref{spectrstart}) is equivalent to (\ref{spectrferm}) provided
the connections are chosen as
\begin{eqnarray}
& A_{\theta}= {\Lambda} +A,  \quad A_{\overline \theta}=
{\overline {\Lambda}} + {\overline A},&\nonumber\\&
{\Lambda}=\left(\begin{array}{ccccc}
0 &   -1 &  0 &  0 &  0 \\
0 &   0 &  0 &  0 &  0 \\
0 &   0 &  0 &  0 &  0 \\
0 &   0 &  0 &  0 &  -1 \\
0 &   0 &  0 &  0 &  0
\end{array}\right),  \quad
{\overline {\Lambda}}=
\left(\begin{array}{ccccc}
0 &   0    &  0 &  -1 &  0 \\
{\lambda}  &  0 &   0 &  0 &  1 \\
0 &   0 &  0 &  0 &  0 \\
0 &   0 &  0 &  0 &  0 \\
0 &   0 &  0 &  {\lambda} & 0
\end{array}\right),
&\nonumber\\ & A=0,  \quad  {\overline
A}=\left(\begin{array}{ccccc}
0 &   0 &  0 &  0 &  0 \\
0 &   0 &  -F &  0 &  0 \\
{\overline D}~{\overline F} &   0 &  0 &  {\cal I}{\overline F} &  0 \\
0 &   0 &  0 &  0 &  0 \\
F{\cal I}{\overline D}~{\overline F}
 &   0 &  -{\overline D}F  &  F{\overline F}&  0
\end{array}\right) &
\label{laxmatrix}
\end{eqnarray}
where ${\Lambda}$ and ${\overline {\Lambda}}$ are constant
matrices and we have introduced the notation
\begin{eqnarray}
{\cal I}_{ab}:= (-1)^{{\overline d}_a}{\delta}_{ab}.
\label{matrixI}
\end{eqnarray}
Using eqs. (\ref{LN=2}--\ref{laxmatrix}) one can derive the even
Lax operator
\begin{eqnarray}
{\cal L}_z := -({\widehat {\cal L}}~{\overline {\cal L}}+
{\widehat {\overline {\cal L}}}{\cal L})={\partial}-{\lambda}E +
{\cal A}, \quad {\cal L}_z {\Psi}=0
\label{laxz}
\end{eqnarray}
where the transformation ${\cal L}\rightarrow {\widehat {\cal L}}$
simply amounts to a change in the sign of the Grassmann-even
matrix entries in ${\cal L}$ and
\begin{eqnarray}
E=\left(\begin{array}{ccccc}
1 &   0 &  0 &  0 &  0 \\
0 &   1 &  0 &  0 &  0 \\
0 &   0 &  0 &  0 &  0 \\
0 &   0 &  0 &  1 &  0 \\
0 &   0 &  0 &  0 &  1
\end{array}\right), \quad
{\cal A} = \left(\begin{array}{ccccc}
0 &   0 &  F &  0 &  0 \\
0 &   0 &  DF &  0 &  0 \\
-D{\overline D}~{\overline F} &
{\overline D}{\cal I}{\overline F} &
0 &  -D{\cal I}{\overline F} &  -{\overline F} \\
-F{\overline D}{\cal I}{\overline F} &
0 &  {\overline D}F &
-F{\overline F} & 0 \\
-DF{\overline D}{\cal I}{\overline F} & F{\overline D}{\cal
I}{\overline F} &  D{\overline D}F & -DF{\overline F} &
-F{\overline F}
\end{array}\right).
\label{laxmatrixz}
\end{eqnarray}

One important remark is in order: {\it the connection ${\cal A}$
(\ref{laxz}--\ref{laxmatrixz}) does not depend on the spectral
parameter ${\lambda}$, and this property is crucial for the
construction that will be carried out}. We would like to emphasize
that there are infinitely many representations equivalent  to
\p{laxmatrix} which generically do not possess this
property\footnote{In other words, in most cases the spectral
parameter appears in field-dependent terms.}. The representation
\p{laxmatrix} is just adapted to use the approach developed in
\cite{adr,adrz,agnpz} (see also references therein).\\
{\bf Remark:} there is, however, another representation with
properties analogous to the representation just described. We
consider the matrix
\be K=\left(\begin{array}{ccccc} 1 & 0 & 0 & 0 & 0\\
0 & 1 & 0 & 0 & 0\\ F & 0 & 1 & 0 & 0\\ 0 & 0 & 0 & 1 & 0\\
\lambda & 0 & -{\overline F} & 0 & 1\end{array}\right),\ee and
transform the operators in \p{LN=2} and \p{laxz} to
\begin{eqnarray}& {\cal L}'={\widehat K}{\cal L}K^{-1}=D+
\left(\begin{array}{ccccc} 0 & -1 & 0 & 0 & 0\\0 & 0 & 0 & 0 & 0
\\ -D{\overline  F} & -{\cal I}{\overline F} & 0 & 0 & 0
\\ \lambda+F{\overline  F} & 0 & -F & 0 & -1
\\ -(DF){\overline F} & -\lambda & {DF} & 0 & 0\end{array}\right),
&\nonumber\\& {\overline {\cal L}}~'={\widehat K}{\overline {\cal L}}K^{-1} =
{\overline D}+\left(\begin{array}{ccccc} 0 & 0 & 0 & -1 & 0
\\0 & 0 & 0 & 0 & 1\\ 0 & 0 & 0 & 0 & 0
\\ 0 & 0 & 0 & 0 & 0
\\0 & 0 & 0 & 0 & 0\end{array}\right),&\nonumber\\&
{\cal L}_z'={ K}{\cal L}_zK^{-1}=\partial-\lambda E+
\left(\begin{array}{ccccc} -F{\overline F} & 0 & F & 0 & 0\\
-(DF){\overline F} & 0 & DF & 0 & 0\\
{\overline D}D{\overline F} & {\overline D}{\cal I}{\overline F} &
0 & -D{\cal I}{\overline F} & {\overline F}\\
{\overline D}(F{\overline F}) & 0 & {\overline D}F & -F{\overline F} & 0\\
{\overline D}((DF){\overline F}) & 0 & -{\overline D}DF & -(DF){\overline F} & 0
\end{array}\right).\label{newrep}
\end{eqnarray}

All matrix entries in formulae \p{laxmatrix}, (\ref{laxmatrixz})
and \p{newrep} are rectangular or square blocks. For instance, all
$1$'s stand for the $k\times k$ identity matrix. A short
inspection shows that after interchanging the components $\psi_2$
and $\psi_5$ ($\psi_2 \leftrightarrow \psi_5 $) of $\Psi$
\p{spectrferm}, the matrices in ${\cal L}_z$ and ${\cal L}_z'$
have the following block structure
\begin{eqnarray}
\left( \begin{array}{cc}
A   &  B \\
C & D
\end{array} \right):=
\left( \begin{array}{c|c}
even   &  odd \\
{\scriptstyle (2k+n)\times (2k+n)} & {\scriptstyle  (2k+n)\times
(2k+m)} \\ \hline
odd & even   \\
{\scriptstyle (2k+m)\times (2k+n) }& {\scriptstyle (2k+m)\times
(2k+m)}\end{array} \right) \label{statistics}
\end{eqnarray}
and zero supertrace
\begin{eqnarray}
Str \left( \begin{array}{cc}
A   &  B \\
C   &  D
\end{array} \right):= Tr(A) - Tr(D) =0,
\label{str}
\end{eqnarray}
so that they belong to the superalgebra
\begin{eqnarray}
{\cal G}=sl(2k+n\vert 2k+m).
\label{supalgtotal}
\end{eqnarray}
The constant matrix $E$ \p{laxmatrixz} defines the splitting
\begin{eqnarray}
&& \quad \quad \quad \quad \quad \quad
{\cal G}= Ker(ad_E) \oplus Im(ad_E), \quad E^2=E, \nonumber\\
&&Ker(ad_E)=
\left(\begin{array}{ccccc}
{}* &   * &  0 &  * &  * \\
{}* &   * &  0 &  * &  * \\
  0 &   0 &  * &  0 &  0 \\
{}* &   * &  0 &  * &  * \\
{}* &   * &  0 &  * &  *  \end{array}\right),
\quad Im(ad_E)=
\left(\begin{array}{ccccc}
0 &   0 &  * &  0 &  0 \\
0 &   0 &  * &  0 &  0 \\
{}* & * &  0 &  * &  * \\
0 &   0 &  * &  0 &  0 \\
0 &   0 &  * &  0 &  0
\end{array}\right)
\label{supalg}
\end{eqnarray}
which possesses the properties
\begin{eqnarray}
&&[Ker(ad_E),~ Ker(ad_E)\}\in Ker(ad_E), \nonumber\\
&&[Ker(ad_E),~ Im(ad_E)\} \in Im(ad_E), \nonumber\\
&&[Im(ad_E),~ Im(ad_E)\}\in Ker(ad_E), \nonumber\\
&& (ad_E)^2\Big|_{Im(ad_E)}= I\Big|_{Im(ad_E)}
\label{supalg1}
\end{eqnarray}
and
\begin{eqnarray}
Ker(ad_E)= s\Bigl(gl(2k\vert 2k)\oplus gl(n \vert m)\Bigr).
\label{subalgker}
\end{eqnarray}

In what follows we will use the homogeneous gradation of the loop
superalgebra
\begin{eqnarray}
{\cal G}\otimes C[{\lambda},{\lambda}^{-1}]
\label{loop}
\end{eqnarray}
with the grading operator
\begin{eqnarray}
d={\lambda}{\textstyle{\partial\over\partial {\lambda}}}.
\label{grad}
\end{eqnarray}
The matrices ${\lambda}E$ and ${\cal A}$  \p{laxmatrixz} entering
into the even Lax operator ${\cal L}_z$ \p{laxz} belong to the
subspaces with grades $1$ and $0$ respectively
\begin{eqnarray}
[d, {\lambda} E] = {\lambda} E, \quad [d, {\cal A}] =0.
\label{grad1}
\end{eqnarray}

We shall construct a non--local gauge transformation $G$, which
commutes with $E$, $GEG^{-1}=E$ and which is fixed by the
requirement that it transforms ${\cal A}$ in (\ref{laxmatrixz}) to
a connection $\widetilde{\cal A}$  belonging to $Im(ad_E)$.
\begin{eqnarray}
\widetilde{\cal A}=G{\cal A}G^{-1}+G\partial G^{-1},\quad
\widetilde{\cal A}\in  Im(ad_E). \label{supalg2}
\end{eqnarray}
With this aim let us first define a $k\times k$ matrix $g$, which
will be useful in what follows, by the consistent set of equations
\begin{eqnarray}
{\partial}g=-gF{\overline F}, \quad
Dg=-\Bigl({\partial}^{-1}g(DF{\overline F})g^{-1}\Bigr)g, \quad
{\overline D}g=-\Bigl({\partial}^{-1}g({\overline D}F {\overline
F})g^{-1}\Bigr)g. \label{def}
\end{eqnarray}
Hereafter, we also use the notation
\begin{eqnarray}
&&f:={\partial}^{-1}gF{\overline D}{\cal I}{\overline F}, \nonumber\\
&&{\overline Q}:= {\overline D}-g^{-1}f, \nonumber\\
&&{\widehat {\overline Q}}~{\overline F}:= {\overline
D}~{\overline F} +{\cal I}{\overline F}g^{-1}f. \label{notat}
\end{eqnarray}
Then, the relevant gauge transformation turns out to be
\begin{eqnarray}
\Psi \quad \Rightarrow \quad \widetilde \Psi =G \Psi, \quad
G=\left(\begin{array}{ccccc}
1 &   0 &  0 &  0 &  0 \\
0 &   1 &  0 &  0 &  0 \\
0 &   0 &  1 &  0 &  0 \\
-f &   0 &  0 &  g &  0 \\
-Df &   f &  0 &  Dg &  g
\end{array}\right)
\label{transf}
\end{eqnarray}
and the corresponding even and odd matrix Lax operators become
\begin{eqnarray}
&{\widetilde {\cal L}}= G{\cal L} G^{-1}=D + {\Lambda},  \quad
{\widetilde {\overline {\cal L}}}= G{\overline {\cal L}} G^{-1}=
{\overline D} + {\widetilde {\overline {\Lambda}}} + {\widetilde
{\overline A}}, \quad {\widetilde {\cal L}}{\widetilde {\Psi}}
={\widetilde {\overline {\cal L}}}{\widetilde {\Psi}}=0, &\nonumber\\
&{\widetilde {\overline {\Lambda}}}= \left(\begin{array}{ccccc}
0 &   0    &  0 &  0 &  0 \\
{\lambda}  &  0 &   0 &  0 &  0 \\
0 &   0 &  0 &  0 &  0 \\
0 &   0 &  0 &  0 &  0 \\
0 &   0 &  0 &  {\lambda} & 0
\end{array}\right), &\nonumber\\
&{\widetilde {\overline A}}=\left(\begin{array}{ccccc}
-g^{-1}f &   0 &  0 &  -g^{-1} &  0 \\
Dg^{-1}f &   -g^{-1}f  &  -F &  Dg^{-1} &  g^{-1} \\
{\widehat {\overline Q}}~{\overline F} &   0 &  0 &
{\cal I}{\overline F}g^{-1} & 0 \\
g {\overline Q}g^{-1}f  &   0 &  0 &  g{\overline Q}g^{-1} &  0 \\
-Dg {\overline Q}g^{-1}f  & -g {\overline Q}g^{-1}f &
-g{\overline Q}F  &  -Dg{\overline Q}g^{-1}& g{\overline Q}g^{-1}
\end{array}\right)&
\label{laxmatrixnew}
\end{eqnarray}
and
\begin{eqnarray}
&{\widetilde {\cal L}}_z =G{\cal L}_zG^{-1}= {\partial}-{\lambda}E
+ {\widetilde {\cal A}}, \quad
{\widetilde {\cal L}}_z {\widetilde {\Psi}}=0, &\nonumber\\
&{\widetilde {\cal A}} = \left(\begin{array}{ccccc}
0 &   0 &  F &  0 &  0 \\
0 &   0 &  DF &  0 &  0 \\
-D{\widehat {\overline Q}}~{\overline F} &
{\widehat {\overline Q}}{\cal I}{\overline F} &
0 &  -D{\cal I}{\overline F}g^{-1} &  -{\overline F}g^{-1} \\
0 &   0 &  g{\overline Q}F & 0 & 0 \\
0 & 0 &  Dg{\overline Q}F  &  0 &  0
\end{array}\right)\in Im (ad_E),&
\label{laxmatrixztransf}
\end{eqnarray}
respectively.

\subsection{Flows}

Now, following ref. \cite{agnpz} one can define flows of the
hierarchy corresponding to the matrix Lax operator
\p{laxmatrixztransf}
\begin{eqnarray}
D_{X_p} {\widetilde {\cal L}}_z =[ ({ X}^{\widetilde\Theta}_p)_{+}
, {\widetilde {\cal L}}_z ], \quad { X}^{{\widetilde\Theta}}_p
={\widetilde\Theta} {\lambda}^p X {\widetilde\Theta}^{-1}, \quad
X\in Ker(ad_E), \quad  p \in \hbox{\bbd N}^+\label{suplaxrepr}
\end{eqnarray}
where $D_{X_p}$ denote the corresponding evolution derivatives,
${\widetilde\Theta}$ is the dressing matrix defined by
\begin{eqnarray}
{{\widetilde\Theta}}^{-1}\Bigl({\partial}-{\lambda}E + {\widetilde
{\cal A}}\Bigr){\widetilde\Theta}={\partial}-{\lambda}E,
\label{dresop}
\end{eqnarray}
and the subscript $+$ denotes the projection on the positive
homogeneous grading \p{grad}. The algebra of the flows
\p{suplaxrepr} is isomorphic to the superalgebra
\begin{eqnarray}
{\widehat {Ker}}(ad_E):= Ker(ad_E)\otimes P({\lambda})
,\label{suploop}
\end{eqnarray}
where $P(\lambda)$ is the set of polynomials in the spectral
parameter $\lambda$. The isospectral flows
${\textstyle{\partial\over\partial t_p}}$ \p{suplax} of the
hierarchy, forming an abelian algebra \p{alg}, have to be
generated by the central element $X=E$ of the kernel $Ker(ad_E)$
via equations \p{suplaxrepr}. All other flows from the set
\p{suplaxrepr} by construction commute with the isospectral flows
and form their bosonic and fermionic symmetries (for detail, see
\cite{agnpz}). To close this subsection let us only mention that
the subalgebra $ sl(2k\vert 2k)\otimes P({\lambda}) \subset
{\widehat {Ker}}(ad_E)$ of the symmetry algebra \p{suploop}
contains two different odd symmetries which may be seen as
extensions of the $N=2$ supersymmetry algebra. Two possible
choices are obtained from the matrices
\begin{eqnarray}
X^{(1)}_{p \pm}= \left(\begin{array}{ccccc}
0 &   {\lambda}^p &  0 &  0 &  0 \\
\pm {\lambda}^{p+1} &   0 &  0 &  0 &  0 \\
0 &   0 &  0 &  0 &  0 \\
0 &   0 &  0 &  0 &  {\lambda}^p \\
0 &   0 &  0 &   \pm {\lambda}^{p+1} &  0  \end{array}\right),
\quad X^{(2)}_{p \pm}= \left(\begin{array}{ccccc}
0 &   0 &  0 &  {\lambda}^p &  0 \\
0 &   0 &  0 &  0 &  {\lambda}^p \\
0 &   0 &  0 &  0 &  0 \\
\pm {\lambda}^{p+1} &   0 &  0 &  0 &  0 \\
0 &   \pm {\lambda}^{p+1} &  0 &  0 &  0
\end{array}\right).
 \label{n=2supergener}
\end{eqnarray}
satisfying the anticommutation relations
\begin{eqnarray}
\{X^{(i)}_{p \pm},~ X^{(i)}_{k \pm}\}= \pm 2{\lambda}^{p+k+1}E,
\quad \{X^{(i)}_{p -},~ X^{(i)}_{k +}\}= 0, \quad i=1,2
\label{n=2superalg}
\end{eqnarray}
The existence of a similar rich symmetry structure for the
particular case of the reduced $N=2$ unconstrained
$(1|1,0)$--MGNLS hierarchy was observed recently in \cite{ks2}.

\subsection{Recursion operators}

Using the general formula for recurrence relations
\begin{eqnarray}
{\textstyle{\partial\over\partial t_p}}{\widetilde {\cal A}} =
\Bigl({\partial}- ad_{{\widetilde {\cal A}}}~{\partial}^{-1}~
ad_{{\widetilde {\cal A}}}\Bigr)~ ad_E
~{\textstyle{\partial\over\partial t_{p-1}}}{\widetilde {\cal A}}
\label{recmatr}
\end{eqnarray}
derived in \cite{adr,agnpz} for the case of Hermitian symmetric
spaces \p{supalg1}, we obtain the following recurrence relations
for the hierarchy under consideration with the Lax operator
${\widetilde {\cal L}}_z$ \p{laxmatrixztransf}:
\begin{eqnarray}
{\textstyle{\partial\over\partial t_p}}F &=&
{\textstyle{\partial\over\partial t_{p-1}}}F~' +
F{\partial}^{-1}{\textstyle{\partial\over\partial t_{p-1}}}
D{\overline D}~{\overline F}F-
Dg^{-1}\Bigl({\textstyle{\partial\over\partial t_{p-1}}}g\Bigr)
{\overline Q}F
+D\Bigl({\partial}^{-1}
{\textstyle{\partial\over\partial t_{p-1}}}F{\widehat {\overline Q}}
{\cal I}{\overline F}\Bigr)F \nonumber\\
&-&\Bigl({\partial}^{-1}{\textstyle{\partial\over\partial
t_{p-1}}}(DF) {\widehat {\overline Q}}{\cal I}{\overline F}\Bigr)F
-\Bigl({\partial}^{-1}{\textstyle{\partial\over\partial
t_{p-1}}}(DF) {\overline F}g^{-1}\Bigr) g{\overline Q}F,
\nonumber\\
{\textstyle{\partial\over\partial t_p}}({\overline F}g^{-1}) &=&
-{\textstyle{\partial\over\partial t_{p-1}}}({\overline F}g^{-1})~' -
\Bigl({\partial}^{-1}{\textstyle{\partial\over\partial t_{p-1}}}
D{\overline D}~{\overline F}F\Bigr){\overline F}g^{-1}
-(D{\widehat {\overline Q}}~{\overline F})
\Bigl({\textstyle{\partial\over\partial t_{p-1}}}g^{-1}\Bigr)\nonumber\\
&+&({\widehat {\overline Q}}{\cal I}{\overline F})
\Bigl({\partial}^{-1}{\textstyle{\partial\over\partial t_{p-1}}}(DF)
{\overline F}g^{-1}\Bigr) - (D{\cal I}{\overline F}g^{-1})
\Bigl({\partial}^{-1}{\textstyle{\partial\over\partial t_{p-1}}}
g({\overline Q}F){\overline F}g^{-1}\Bigr) \nonumber\\
&-&{\overline F}g^{-1}
\Bigl({\partial}^{-1}{\textstyle{\partial\over\partial t_{p-1}}}(Dg
{\overline Q}F){\overline F}g^{-1}\Bigr)
\label{recFbar}
\end{eqnarray}
where $'$ denotes the derivative with respect to the space
variable $z$.

We have verified explicitly by direct calculations that the first
few bosonic flows generated by eqs. \p{recFbar} with the initial
recursion step
\be
{\textstyle{\partial\over\partial t_1}}F=F~', \quad
{\textstyle{\partial\over\partial t_1}}{\overline F}={\overline F}~'
\ee
reproduce the corresponding isospectral flows
${\textstyle{\partial\over\partial t_p}}$ of the $N=2$ supersymmetric
unconstrained $(k|n,m)$--MGNLS hierarchy resulting from the
pseudo--differential Lax--pair representation \p{suplax}.

\subsection{Supersymmetry and locality}

Although at the component level, the non-zero matrix entries in
the connexion ${\widetilde {\cal A}}$ in \p{laxmatrixztransf} are
all independent, this is not so at the superfield level. The
connexion satisfies constraints, which may be written as \be
[{\widetilde {\cal L}}, {\widetilde {\cal L}}_z\}={\widetilde
{\cal L}}{\widetilde {\cal L}}_z-{\widehat {\widetilde {\cal
L}}}_z{\widetilde {\cal L}}=0,\quad [{\widetilde {\overline {\cal
L}}}, {\widetilde {\cal L}}_z\}={\widetilde {\overline {\cal
L}}}{\widetilde {\cal L}}_z-{\widehat {\widetilde {\cal
L}}}_z{\widetilde {\overline {\cal L}}}=0.\label{constt}\ee If
these constraints are respected by the flows, then the flows are
consistent with supersymmetry. In fact, only the first of these
constraints is easily shown to be respected by the isospectral
flows. Using the dressing equation \p{dresop}, we rewrite this
constraint as \be
[{\widehat{\widetilde\Theta}}^{-1}(D+\Lambda){\widetilde\Theta},
\partial-\lambda E\}.\ee
Considering this equation at each homogeneous gradation, it is
 easy to show that it leads to
\be{\widehat{\widetilde\Theta}}^{-1}(D+\Lambda){\widetilde\Theta}=D+\Lambda.\ee
It is then clear that the matrix $E^{{\widetilde\Theta}}_p
={\widetilde\Theta} {\lambda}^p E {\widetilde\Theta}^{-1}$
commutes with the operator $D+\Lambda$. Since this last operator
respects the homogeneous gradation, we end up with the equation
\be [D+\Lambda,(E^{{\widetilde\Theta}}_p)_{+}\}=0,\ee which shows
that the isospectral flows respect the first of constraints
\p{constt}. We conjecture that the second of these constraints is
also preserved, although we could not show it.

Let us discuss shortly the locality of the isospectral flows
\p{suplaxrepr} with $X=E$. When rewritten in terms of the local
operator ${\cal L}_z$ in \p{laxz}, they become
\begin{eqnarray}
D_{E_p} { {\cal L}}_z =[ ({ E}^{\Theta}_p)_{+}-G^{-1}D_{E_p}G ,
{\cal L}_z ], \quad { E}^{{\Theta}}_p ={\Theta}
{\lambda}^p E {\Theta}^{-1},  \quad p \in \hbox{\bbd
N}^+,\label{locallaxrepr}
\end{eqnarray}
where the matrix $\Theta$ is obtained from dressing the operator
${\cal L}_z$
\begin{eqnarray}
{{\Theta}}^{-1}\Bigl({\partial}-{\lambda}E + { {\cal
A}}\Bigr){\Theta}={\partial}-{\lambda}E.
\label{dresopnew}
\end{eqnarray}
It is known that the matrix $({ E}^{\Theta}_p)_{+}$ is a local
functional in the fields and their derivatives. Moreover, from the
form of $G$ in \p{transf} one can show that the second term
$-G^{-1}D_{E_p}G$ of the Lax representation \p{locallaxrepr} does
not contribute to the field equation of $F$, which is thus local.
This is not so, however, for $\overline F$.

We conjecture that in order to demonstrate completely the
supersymmetry  and locality of the isospectral flows, one should
make use of the second representation introduced in \p{newrep}.
This point is still under investigation.

\section{Conclusion}

In this letter we have constructed a $sl(2k+n\vert
2k+m)$--super--algebraic formulation
(\ref{laxmatrixnew}--\ref{laxmatrixztransf}) of the $N=2$
supersymmetric unconstrained $(k|n,m)$--MGNLS hierarchies in $N=2$
superspace. Then we have derived the superalgebra
$s\Bigl(gl(2k\vert 2k)\oplus gl(n \vert m)\Bigr)\otimes
P({\lambda})$ \p{suploop} of their fermionic and bosonic
symmetries \p{suplaxrepr}. We have observed that this symmetry
superalgebra contains many odd flows, some of them generalizing
the $N=2$ supersymmetry algebra. Finally we have constructed the
recursion relations \p{recFbar} for these hierarchies.

Let us finish this Letter with a few questions for the future. It
is easily seen that the connection ${\widetilde {\cal A}}$
entering into the Lax operator ${\widetilde {\cal L}}_z$
\p{laxmatrixztransf} is nonlocal. Moreover, its $N=2$ superfield
entries are not independent\footnote{Though, at the component
level, the entries in ${\widetilde {\cal A}}$ \p{laxmatrixztransf}
are independent, and fill up the image $Im(ad_E)$.} quantities,
i.e. they are subjected to constraints. Why in this case do
isospectral matrix flows \p{suplaxrepr} be local, as it is
obviously the case in their original pseudo--differential
representation \p{suplax}? Why are they supersymmetric, or in
other words, why do these flows preserve the above--mentioned
constraints? Finally, how can one see in general that these flows
coincide with the original flows \p{suplax} we started with. These
questions are clarified only partly in the present Letter, and we
hope to discuss them in more detail elsewhere.

{}~

{}~

\noindent{\bf Acknowledgments.} A.S. would like to thank the
Laboratoire de Physique de l'ENS Lyon for the hospitality during
the course of this work. This work was partially supported by the
PICS Project No. 593, RFBR-CNRS Grant No. 01-02-22005, Nato Grant
No. PST.CLG 974874, RFBR-DFG Grant No. 02-02-04002 and
DFG Grant 436 RUS 113/669.


\begin{thebibliography}{**}
\bibitem{ks1}
A.S. Sorin and P.H.M. Kersten, {\it The N=2 supersymmetric
unconstrained matrix GNLS hierarchies}, nlin.SI/0201026, Lett.
Math. Phys. (in press).
\bibitem{bks1}
L. Bonora, S. Krivonos and A. Sorin, {\it The $N=2$ supersymmetric
matrix GNLS hierarchies}, Lett. Math. Phys. {\bf 45} (1998) 63,
solv-int/9711009.
\bibitem{bks2}
L. Bonora, S. Krivonos and A. Sorin, {\it Coset approach to the
$N=2$ supersymmetric matrix GNLS hierarchies}, Phys. Lett. {\bf
A240} (1998) 201, solv-int/9711012.
\bibitem{bks}
L. Bonora, S. Krivonos and A. Sorin, {\it Towards the construction
of $N=2$ supersymmetric integrable hierarchies}, Nucl. Phys. {\bf
B477} (1996) 835, hep-th/9604165.
\bibitem{bs}
L. Bonora and A. Sorin, {\it The Hamiltonian structure of the N=2
supersymmetric GNLS hierarchy}, Phys. Lett. {\bf B407} (1997) 131,
hep-th/9704130.
\bibitem{pop}
Z. Popowicz, {\it The extended supersymmetrization of the
multicomponent Kadomtsev-Petviashvili hierarchy}, J. Phys. {\bf
A29} (1996) 1281, hep-th/9510185.
\bibitem{fk}
A.P. Fordy and P.P. Kulish, {\it Nonlinear Schr\"{o}dinger
equations and simple Lie algebras}, Commun. Math. Phys. {\bf 89}
(1983) 427.
\bibitem{ks2}
P.H.M. Kersten and A.S. Sorin, {\it Bi-Hamiltonian structure of
the $N=2$ supersymmetric ${\alpha}=1$ KdV hierarchy},
nlin.SI/0201061, Phys. Lett. {\bf A} (in press).
\bibitem{dm1}
F. Delduc and M. Magro,
{\it Gauge invariant formulation of $N=2$ Toda and KdV systems in extended
superspace}, J.Phys. {\bf A29} (1996) 4987, hep-th/9512220.
\bibitem{dm2}
F. Delduc and M. Magro,
{\it $N=2$ chiral WZNW model in superspace},
Int.J.Mod.Phys. {\bf A11} (1996) 4815, hep-th/9512221.
\bibitem{dfg}
F. Delduc, L. Feher and L. Gallot, {\it Nonstandard
Drinfeld-Sokolov reduction}, J. Phys. A : Math. Gen. {\bf 31}
(1998) 5545, solv-int/9708002.
\bibitem{dg}
F. Delduc and L. Gallot, {\it Supersymmetric Drinfeld-Sokolov
reduction}, J. Math. Phys. {\bf 39} (1998) 4729, solv-int/9802013.
\bibitem{mm}
J.O. Madsen and J.L. Miramontes,
{\it Non-local conservation laws and flow equations for
supersymmetric integrable hierarchies},
Commun.Math.Phys. {\bf 217} (2001) 249, hep-th/9905103.
\bibitem{adr}
H. Aratyn, A. Das and C. Rasinariu, {\it Zero Curvature Formalism
for Supersymmetric Integrable Hierarchies in Superspace}, Mod.
Phys. Lett. {\bf A12} (1997) 2623, hep-th/9704119.
\bibitem{adrz}
H. Aratyn and A. Das, {\it The sAKNS hierarchy}, Mod. Phys. Lett.
{\bf13} (1998) 1185, solv-int/9710026; \\H. Aratyn, A. Das, C.
Rasinariu and A.H. Zimerman, {\it Zero curvature formalism in
superspace}, in {\it Supersymmetry and Integrable Models},
Proceedings of the UIC-Theory Workshop, June 1997, H. Aratyn et al
(Eds) Springer-Verlag, 1998 (Lecture Notes in Physics 502).
\bibitem{agnpz}
H. Aratyn, J.F. Gomes, E. Nissimov, S. Pacheva and A.H. Zimerman,
{\it Symmetry Flows, Conservation Laws and Dressing Approach to
the Integrable Models}, in {\it Integrable Hierarchies and Modern
Physical Theories}, Eds. H. Aratyn and A.S. Sorin, Kluwer Acad.
Publ., Dordrecht/Boston/London, 2001, pg. 243, nlin.SI/0012042.

\end{thebibliography}
\end{document}